# Supergiant Barocaloric Effects in Acetoxy Silicone Rubber over a Wide Temperature Range: Great Potential for Solid-state Cooling

William Imamura[a,], Érik Oda Usuda[,b], Lucas Soares Paixão[c], Nicolau Molina Bom[d,e], Angelo Marcio Gomes[f], Alexandre Magnus Gomes Carvalho[g *]

[a] *Faculdade de Engenharia Mecânica, UNICAMP, CEP 13083-860, Campinas, SP, Brazil.*
[b] *Universidade Federal de São Paulo, UNIFESP, CEP 09913-030, Diadema, SP, Brazil.*
[c] *Instituto de Física Gleb Wataghin, UNICAMP, CEP 13083-859, Campinas, SP, Brazil.*
[d] *Catalan Institute of Nanoscience and Nanotechnology, ICN2, 08860 Belaterra, Barcelona, Spain*
[e] *The Institute of Photonic Sciences, ICFO, 08193 Castelldefels, Barcelona, Spain*
[f] *Instituto de Física, UFRJ, CEP 21941-972, Rio de Janeiro, RJ, Brazil.*
[g] *Departamento de Engenharia Mecânica, UEM, CEP 87020-900, Maringá, PR, Brazil.*
*E-mail: amgcarvalho.pem.uem@gmail.com

ABSTRACT

Solid-state cooling based on caloric effects is considered a viable alternative to replace the conventional vapor-compression refrigeration systems. Regarding barocaloric materials, recent results show that elastomers are promising candidates for cooling applications around room-temperature. In the present paper, we report supergiant barocaloric effects observed in acetoxy silicone rubber – a very popular, low-cost and environmentally friendly elastomer. Huge values of adiabatic temperature change and reversible isothermal entropy change were obtained upon moderate applied pressures and relatively low strains. These huge barocaloric changes are associated both to the polymer chains rearrangements induced by confined compression and to the first-order structural transition. The results are comparable to the best barocaloric materials reported so far, opening encouraging prospects for the application of elastomers in near future solid-state cooling devices.

## 1. INTRODUCTION

The current priorities in sustainability and energy efficiency leads to the study and development of new refrigeration technologies. In this context, solid-state cooling based on caloric effects (also called *i*-caloric effects[1] has shown to be a promising alternative to the conventional vapor-compression systems. The caloric effects can be described by an isothermal entropy change ($\Delta S_T$) and an adiabatic temperature change ($\Delta T_S$), both manifested when an external field is applied on a material. Depending on the nature of this external field (magnetic field, electric field or stress field), the caloric effects can be categorized as magnetocaloric, electrocaloric effect and mechanocaloric effect ($\sigma$-CE). The latter can still be divided in elastocaloric effect, driven by uniaxial stress; and barocaloric effect, driven by isotropic stress variations.

The first caloric effect reported was the elastocaloric effect in natural rubber (NR), observed by John Gough in the beginning of the nineteenth century.[2] Nevertheless, the research into caloric effects has strengthened only in the last two decades, due to the discoveries of giant magnetocaloric effect in $Gd_5Si_2Ge_2$ compound[3] and giant electrocaloric effect in $PbZr_{0.95}Ti_{0.05}O_3$ thin film.[4] Even though $\sigma$-CE is the least researched among caloric effects, interesting results were reported for vulcanized natural rubber (VNR) and other synthetic elastomers already in the 1940 decade.[5],[6] Shape-memory alloys also exhibit promising $\sigma$-CE around room temperature.[7]–[11] More recently, giant barocaloric effects were reported in $(NH_4)_2SO_4$ and molecular crystal below room temperature.[12],[13] Another promising class of material is plastic crystals, which demonstrated colossal barocaloric effects near room temperature.[14],[15] Regarding elastomers, the number of studies reporting large $\sigma$-CE values is gradually growing in the last years.[16]–[23] On the contrary of what is observed in shape memory alloys and ionic salts, elastomers show significant elastocaloric and barocaloric effects even in absence of phase transitions. This behavior is assigned to the rearrangement of the polymer chains induced by the application of mechanical stress.[16]

In this context, the silicone rubber can be considered a potential mechanocaloric material, since it is generally elastomeric and may present favorable structural transitions. The popular term "silicone" includes any organosilicon compound containing at least one pair of silicon atoms linked by an oxygen atom (Si–O–Si). Properly speaking, the correct terminology for these compounds is "polysiloxane", whose formula is $(RR'SiO)_n$ – where R and R' are alkyl groups.[24]

Among the polysiloxanes, there is a group named room-temperature vulcanizing silicone rubbers (RTV-SR), which consist in polydimethylsiloxane, curing agent, fillers and additives. Acetoxy silicone rubber (ASR) is a type of RTV-SR, which releases acetic acid during vulcanization. Recently, our preprint results [25] motivated studies on ASR as a viable alternative for using as a refrigerant in solid-state devices based in barocaloric effects.[26]–[28] In the present paper, we systematically investigate the barocaloric effect in ASR around room temperature. The $\Delta S_T$ was evaluated following an indirect method using a Maxwell relation, and $\Delta T_S$ was directly measured. The results allowed us to establish a link between the crystalline–amorphous transition and the very large barocaloric changes observed in ASR.

## 2. MATERIALS AND METHODS

ASR samples were prepared by filling cylindrical plaster molds with a commercial silicone sealant resin (TYTAN Professional®) and were left to dry for approximately 24 hours, to complete the room-temperature vulcanization process. For pressures up to 173 MPa, 12-mm-diameter samples were used; above 173 MPa, 8-mm-diameter samples were used. Both samples had length of ~20 mm and the average density was estimated by a pycnometer in 960 kg m$^{-3}$.

The experiments with pressure were performed in the experimental setup detailed elsewhere; the isostatic condition for barocaloric measurements was also discussed.[20],[21],[23],[29] Figure S1, in *ESI*, presents a schematic representation of the apparatus used in the barocaloric experiments. Two versions of the pressure chamber are available, consisting of carbon-steel cylinders with an 8 mm or 12 mm bore. The maximum pressure attainable with each chamber is 173 MPa (12 mm) and 390 MPa (8 mm). Samples are prepared in the proper shape to perfectly fit the chamber. A piston is the only movable part, responsible for compressing the sample. The bottom closure of the chamber has a narrow hole allowing a thermocouple to be placed inside the sample, monitoring its temperature. Another thermocouple, placed inside the chamber wall, monitors the chamber temperature, and its reading is used as a feedback to the temperature control system. A copper coil guides fluids (water or liquid nitrogen) around the chamber, and together with the heating elements, is responsible for controlling the chamber temperature. The chamber is placed on a load gauge and a hydraulic press actuates the piston. The pressure applied to the sample is directly given by the force measured by the load gauge divided by the piston cross-section area. A length gauge monitoring the displacement of the piston retrieves information about the deformation of the sample.

$\Delta T_S$ as a function of temperature (223–333 K) was directly measured using a quasi-adiabatic compression/decompression process. During such process, pressure varies quickly enough to preclude significant heat exchange between the sample and its surroundings. Pressure changes between the minimum and the maximum pressures ($p_1$ and $p_2$, respectively) were in the range of 26.0–390 MPa; compression corresponds to the process $p_1 \rightarrow p_2$, while decompression corresponds to the process $p_2 \rightarrow p_1$.

Strain *vs.* temperature curves were measured at different constant pressures (0.9–332 MPa), varying the temperature in a rate of ~3 K/min for cooling and heating processes, within the 213–333 K temperature range. Strain ($\varepsilon$) is defined as $\varepsilon(p,T) \equiv (l_{p,T}-l_0)/l_0$, where $l_{p,T}$ is the final length of the sample at pressure $p$, for each temperature $T$, and $l_0$ is its initial length measured at ambient pressure ($p_0$) and room temperature ($T_0$ = 293 K). Since the cross-sectional area ($A$) of the samples does not vary in confined pressure, the volume change ($\Delta V$) is directly proportional to the strain because $\varepsilon(p,T) \equiv (Al_{p,T} - Al_0)/Al_0 = \Delta V/V_0$. These strain *vs.* temperature curves up to 173(3) MPa were used in the calculation of $\Delta S_T$ *vs.* $T$.

X-ray diffraction (XRD) patterns were measured at the XRD1 beamline,[30],[31] at the Brazilian Synchrotron Light Laboratory (LNLS). The beam energy used at XRD1 was 12 keV, and the samples were cooled down by a Cryojet5 (Oxford Instruments), within the 300–100 K temperature range at ambient pressure.

Fourier transform infrared spectroscopy (FTIR) was performed by a spectrometer from PerkinElmer® (model Spectrum Two), in the spectral range from 4000 to 450 cm$^{-1}$, with spectral resolution of 1 cm$^{-1}$ (Figure S2, *ESI*).

## 3. RESULTS AND DISCUSSION

In each $\varepsilon$ *vs.* $T$ curves on heating process (Figure 1a), one can see a narrow region where the derivative abruptly increases, and the transition temperature ($T_{TR}$) shifts toward higher temperatures for larger applied pressures. This behavior, in addition to the hysteresis shown in Figure 1b, strongly indicates a first-order transition. In fact, as observed in the XRD patterns (Figure 1c), there is an amount of amorphous phase that becomes crystalline phase when the temperature decreases. To verify the influence of the pressure on $T_{TR}$, we calculated each $T_{TR}$ from a local maximum in d$\varepsilon$/d$T$ *vs.* $T$ curves. Considering a linear fit, we found d$T_{TR}$/d$p$ = 0.27(1) K MPa$^{-1}$ on heating and 0.23(1) K MPa$^{-1}$ on cooling (Figure 1d).

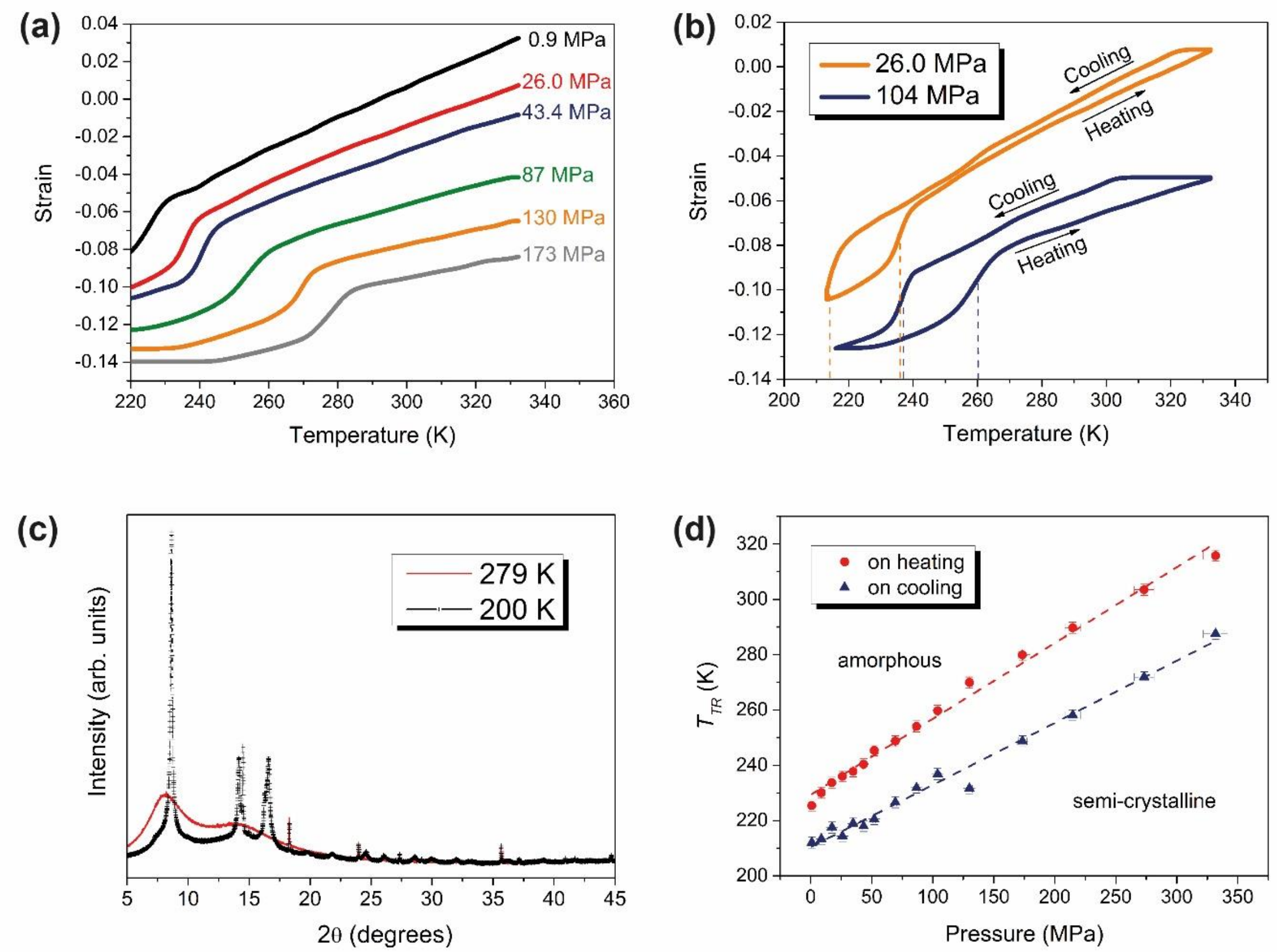

**Figure 1.** (Color online) Crystalline–amorphous transition of ASR. (a) Strain *vs.* temperature curves at constant pressures of 0.9(1), 26,0(5), 43.4(9), 87(2), 130(3) and 173(3) MPa measured on heating. (b) Strain *vs.* temperature curves obtained during the cooling and heating processes, at 26.0(5) and 104(2) MPa; the vertical dotted lines indicate the transition temperatures ($T_{TR}$), on heating and on cooling processes. (c) X-ray diffraction patterns corresponding to measurements at 279 K (amorphous phase) and 200 K (crystalline phase), at ambient pressure; see Figure S3 and Table S2 in *ESI* for crystallographic information. (d) $T$ *vs.* $p$ phase diagram: Transition temperature on heating and on cooling processes *vs.* applied pressure; solid symbols are experimental data, and dash lines are the linear fits to obtain d$T_{TR}$/d$p$; we estimate errors of ±2% for pressures up to 173 MPa and ±3% above 173 MPa, and ±2 K for $T_{TR}$.

Temperature as a function of time for ASR (see an example in Figure 2a) was measured at the pressure range of 26.0(5)–390(12) MPa and different initial temperatures (223–333 K). The observed behavior is reversible in a large temperature range (*i.e.*, the adiabatic temperature change during compression and decompression is similar). The asymmetry found at the highest pressure in Figure 2a (the absolute values of temperature change in decompression is higher than in compression for 390 MPa) is related to the partial amorphous-crystalline phase transition. In Figure 2b, the $\Delta T_S$ corresponding to the decompression process is displayed. We observe a maximum barocaloric effect of 41.1 K, at ~298 K, for $|\Delta p|$ = 390 MPa. This $\Delta T_S$ value, which we classify as supergiant ($|\Delta T_S| \geq 30$ K), is significantly higher than those reported for any barocaloric materials around room temperature (*e.g.*, VNR[21] at ~315 K presents $|\Delta T_S|$ = 24.9 K for $|\Delta p|$ = 390 MPa; PDMS[20] at ~283 K presents $|\Delta T_S|$ = 28.5 K for $|\Delta p|$ = 390 MPa). It is easy to see that a $|\Delta T_S|$ maximum appears for $|\Delta p|$ = 173 MPa, and this maximum shifts to higher temperatures when the pressure increases. Considering a linear rate for the temperatures at the maximum $\Delta T$ ($T_{m,\Delta T}$) with pressure, we have $\mathrm{d}T_{m,\Delta T}/\mathrm{d}p$ = 0.22(8) K MPa$^{-1}$. The outstanding $\Delta T_S$ values registered in ASR can be understood as a combination of the structural changes associated to the crystalline–amorphous transition and polymer chains rearrangements unrelated to phase transitions. Above $T_{m,\Delta T}$, the contribution to the barocaloric effect comes entirely from the amorphous phase. If we compare the $|\Delta T_S|$ values at $T_{m,\Delta T}$ with the $|\Delta T_S|$ values above $T_{m,\Delta T}$, an increase of 33, 34 and 47% is obtained for 173, 273 and 390 MPa, respectively. In addition, $|\varepsilon|$ is less than 25% up to $|\Delta p|$ = 390 MPa.

The $\varepsilon$ *vs*. $T$ curves on heating and on cooling processes were used to calculate the $\Delta S_T$ as a function of temperature (Figure 3a-b), according to Eq. (1), derived from Maxwell's relation:[19]–[21]

$$\Delta S_T(T, \Delta p) = -\rho_0^{-1} \int_{p_1}^{p_2} \left(\frac{\partial \varepsilon_V}{\partial T}\right)_p dp, \qquad (1)$$

where $\rho_0$ is the density at $p_0$ and $T_0$, and $p_1 \approx p_0$.

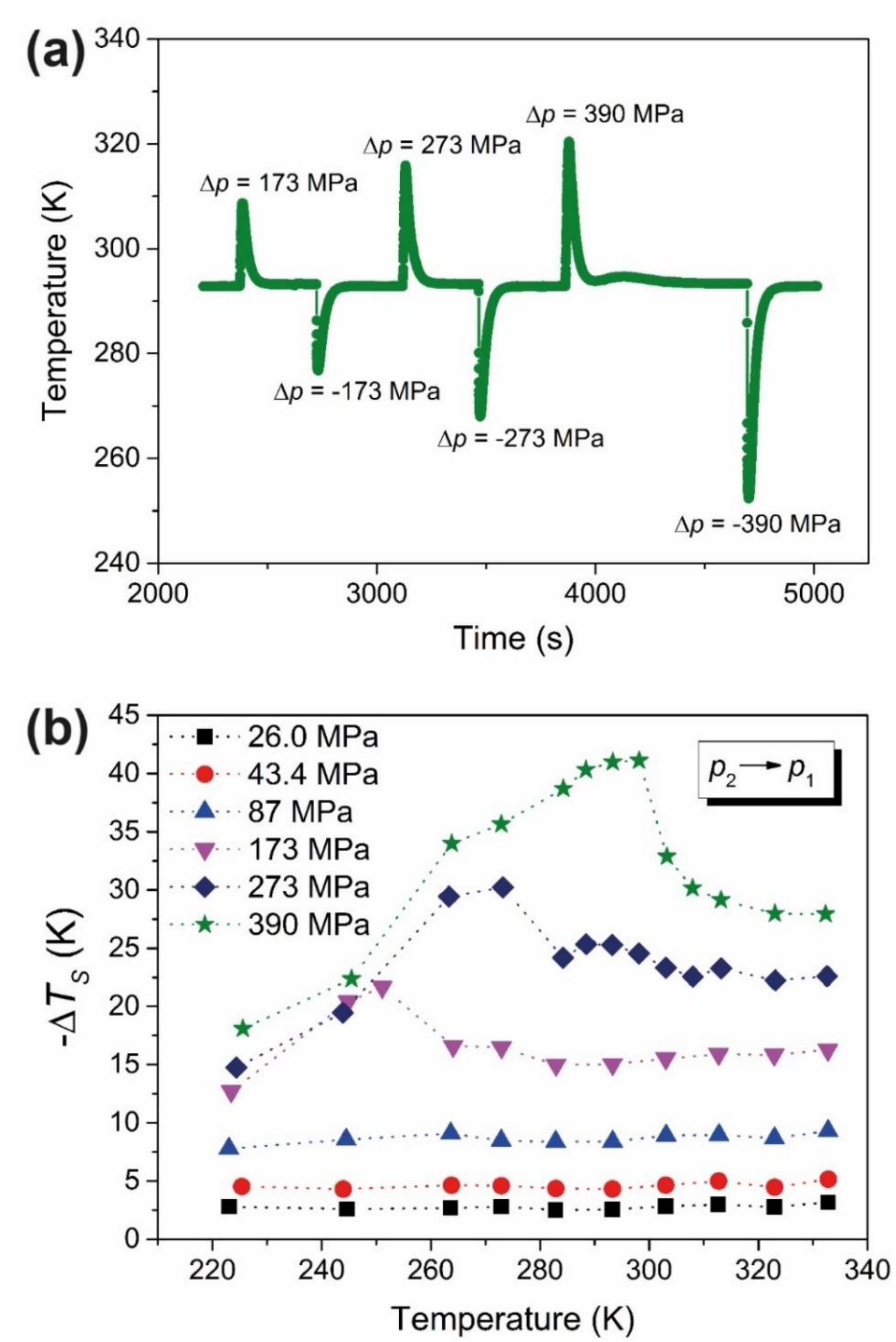

**Figure 2.** (Color online) Adiabatic temperature change. (a) Temperature *vs.* time for ASR during a barocaloric cycle at initial temperature of 293 K and pressure changes of 173(3), 273(8) and 390(12) MPa. (b) Adiabatic temperature change *vs.* initial temperature for decompression process at different pressure variations; the dotted lines connecting the symbols are guides for the eyes; we estimate an error of ±2% for pressures up to 173 MPa and ±3% above 173 MPa, and an asymmetric error of +4% for $|\Delta T_S|$; $p_1$ and $p_2$ are the minimum and the maximum pressures, respectively.

Taking $\Delta S_T$ on heating process as example, very large $|\Delta S_T|$ values were obtained, whose maxima are around 65(13), 100(20), 160(32) and 210(42) J kg$^{-1}$ K$^{-1}$ for pressure changes of 26.0(5), 43.4(9), 87(2) and 173(3) MPa, respectively. The corresponding normalized entropy changes ($|\Delta S_T/\Delta p|$) are 2.5(6), 2.4(5), 1.8(4) and 1.2(3) kJ kg$^{-1}$ K$^{-1}$ GPa$^{-1}$, which are much larger than those reported for PDMS[20] and VNR.[19],[21] $|\Delta S_T|$ values increase more drastically due to the crystallization than the $\Delta T_S$ discussed above. For instance, the $|\Delta S_T|$ value at the maximum for $|\Delta p|$ = 26.0 MPa are more than twice the value in the amorphous region. For $|\Delta p|$ = 43.4 MPa, we observe giant $\Delta S_T$ values at the amorphous region and supergiant $\Delta S_T$ ($|\Delta S_T| \geq 90$ J kg$^{-1}$ K$^{-1}$) at the peak, due to the structural transition. We observed that the temperature at the maximum $|\Delta S_T|$ ($T_{m,\Delta S}$) shifts to higher values when the pressure increases, increasing at a rate of $\mathrm{d}T_{m,\Delta S}/\mathrm{d}p$ = 0.26(3) K MPa$^{-1}$, analogously to the maxima in $-\Delta T_S$ *vs*. $T$ curves (Figure 2b, $\mathrm{d}T_{m,\Delta T}/\mathrm{d}p$ = 0.22(8) K MPa$^{-1}$). Thus, $\Delta S_T$ and $\Delta T_S$ maxima shifts similarly with pressure (taking the errors into account), although obtained from different processes. As expected, $\mathrm{d}T_{m,\Delta S}/\mathrm{d}p \cong \mathrm{d}T_{TR}/\mathrm{d}p$, since $\Delta S_T$ curves from Figure 3a are obtained from $\varepsilon$ *vs*. $T$ data on heating process. Similar behaviors can be observed taking $\Delta S_T$ on cooling process (Figure 3b). Moreover, $\Delta S_T$ on heating and on cooling show irreversibility due to the hysteresis (*i.e.*, the difference between $\Delta S_T$ from heating and cooling processes cannot be negligible). In order to eliminate the irreversible contribution, the reversible $\Delta S_T$ as a function of temperature was estimated from the overlapping $\Delta S_T$ on heating and on cooling processes (Figure 3c). We can see that the reversible $|\Delta S_T|$ is still supergiant (maximum of 182(46) J kg$^{-1}$ K$^{-1}$ for $|\Delta p|$ = 173 MPa). This reversibility is essentially due to the amorphous phase. Regarding the peaks observed on Figure 3, we must note that local fluctuations occur across the entire temperature and pressure ranges; such fluctuations are artifacts generated by the experimental and numerical precision of the measurements and data analysis procedures.

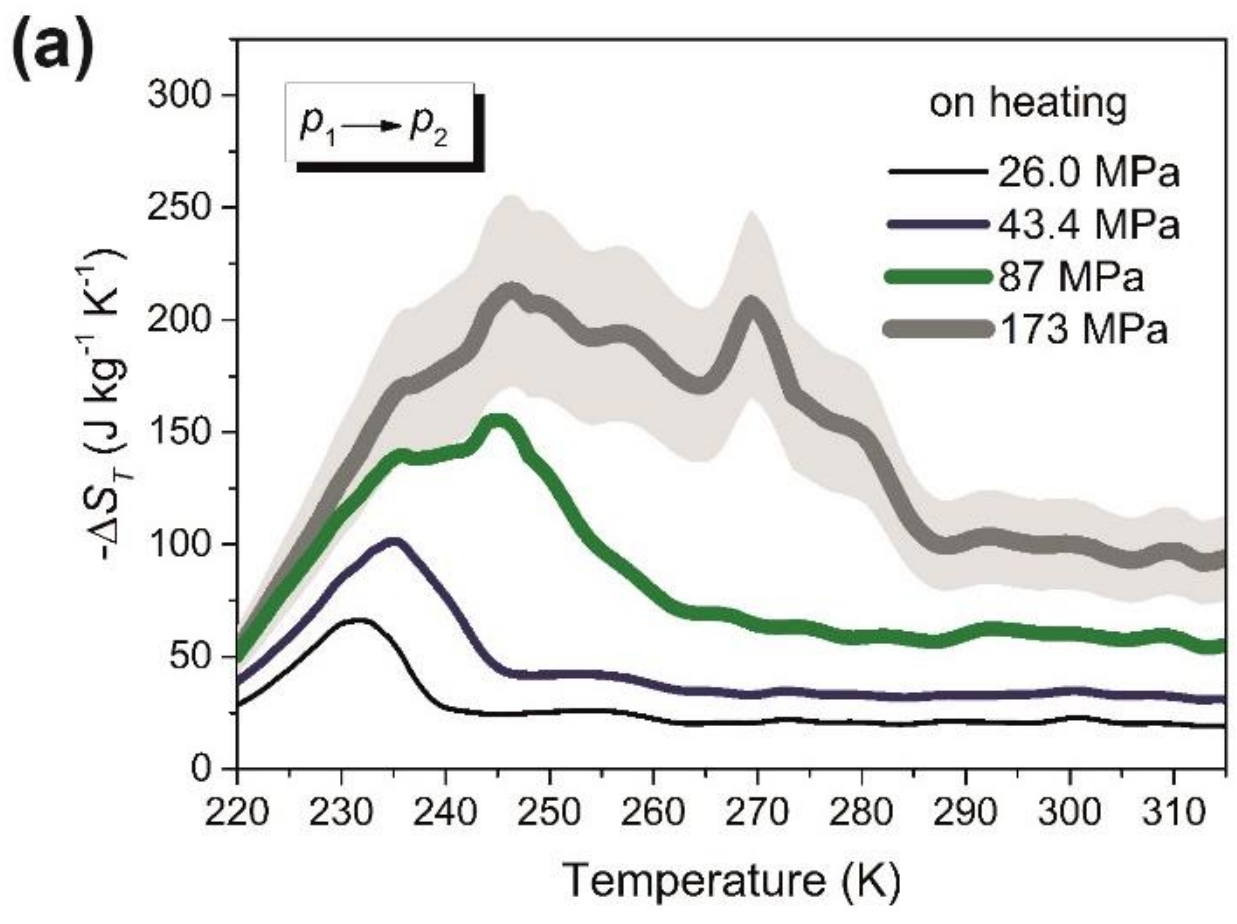

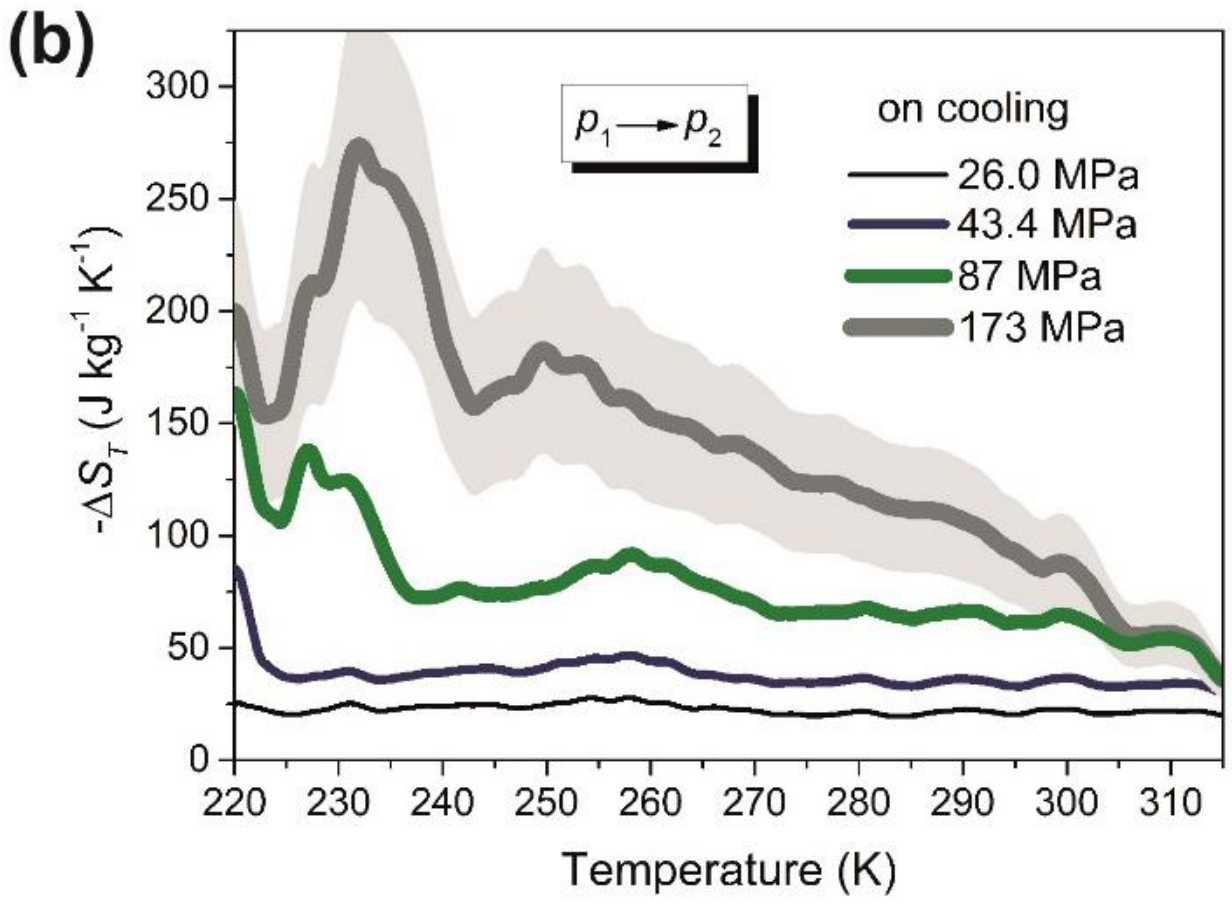

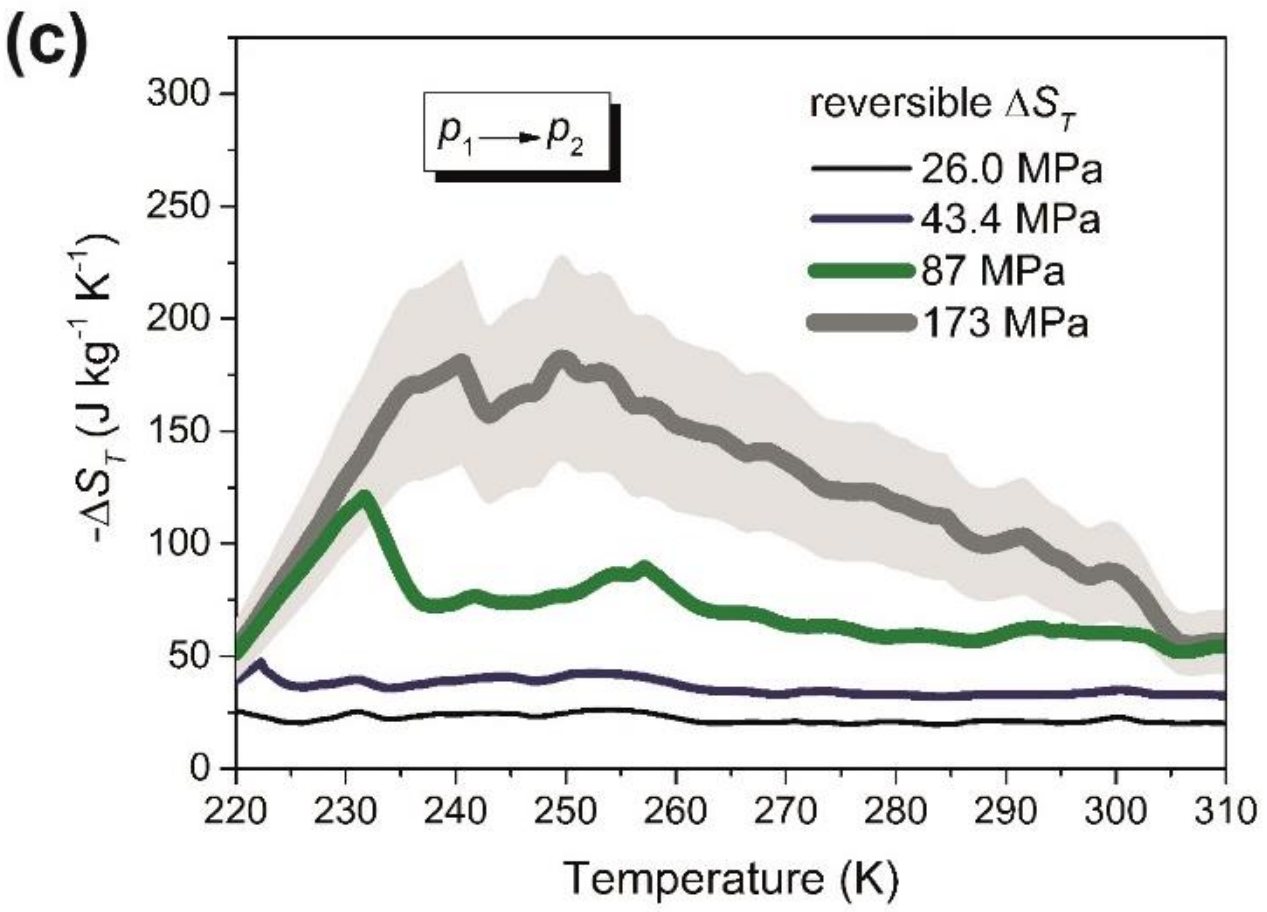

**Figure 3.** (Color online) Isothermal entropy change. (a) Isothermal entropy change as a function of temperature for $|\Delta p|$ = 26.0(5), 43.4(9), 87(2) and 173(3) MPa, obtained from $\varepsilon$ *vs*. $T$ data measured on heating process using Eq. (1); the errors were estimated in ±20%. (b) Isothermal entropy change as a function of temperature for $|\Delta p|$ = 26.0(5), 43.4(9), 87(2) and 173(3) MPa, obtained from $\varepsilon$ *vs*. $T$ data measured on cooling process using Eq. (1); the errors were estimated in ±25%. (c) Reversible isothermal

entropy change as a function of temperature for $|\Delta p|$ = 26.0(5), 43.4(9), 87(2) and 173(3) MPa, obtained from overlapping Figure 3a and Figure 3b; the errors were estimated in ±25%. The errors were only shown for 173 MPa (gray-filled area) for the sake of clarity. $p_1$ and $p_2$ are the minimum and the maximum pressures, respectively.

Finally, we compared the barocaloric properties of ASR around room temperature with promising barocaloric materials in literature.[9],[11],[19]–[21],[32]–[35] The *normalized adiabatic temperature change* ($|\Delta T_S/\Delta p|$) as a function of $|\Delta T_S|$ is plotted in Figure 4a, where simultaneously larger $|\Delta T_S|$ and $|\Delta T_S/\Delta p|$ values indicate higher potential for barocaloric cooling applications. $|\Delta T_S/\Delta p|$ reaches a huge maximum value of ~120 K GPa$^{-1}$ for $|\Delta T_S|$ = 20.4 K. It is noteworthy the results for ASR exceed those of any other barocaloric material in a broad temperature range. We also calculated the *normalized refrigerant capacity* (*NRC*) as a function of the temperature difference between hot reservoir and cold reservoir ($\Delta T_{h\text{-}c} \equiv T_{hot} - T_{cold}$), following the equation:[20],[21]

$$NRC(\Delta T_{h-c}, \Delta p) = \left|\frac{1}{\Delta p}\int_{T_{cold}}^{T_{hot}} \Delta S_T(T, \Delta p)dT\right|, \qquad (2)$$

as shown in Figure 4b. For ASR, we fixed the hot reservoir at 300 K, $|\Delta p|$ = 173 MPa and we used $\Delta S_T$ data from Figure 3a and reversible $\Delta S_T$ data from Figure 3c. Again, the *NRC* values of ASR surpasses all barocaloric materials in the full $\Delta T_{h\text{-}c}$ range, and this difference increases as a function of $\Delta T_{h\text{-}c}$, reaching ~15 kJ kg$^{-1}$ GPa$^{-1}$ for $\Delta T_{h\text{-}c}$ = 25 K. Moreover, the curve keeps a clear trend to increase, following a distinct behavior to that observed for the other non-elastomeric barocaloric materials in the comparison. Also, the relative cooling power ($RCP \equiv |\Delta S_{max} \times \delta T_{FWHM}|$, where $\Delta S_{max}$ is the maximum entropy change and $\delta T_{FWHM}$ is the full width of the entropy change peak at half maximum) for ASR taking reversible $\Delta S_T$ is 13(3) kJ kg$^{-1}$ for $|\Delta p|$ = 173(3) MPa; this *RCP* value is so huge that exceed all values reported for barocaloric materials by at least one order of magnitude.

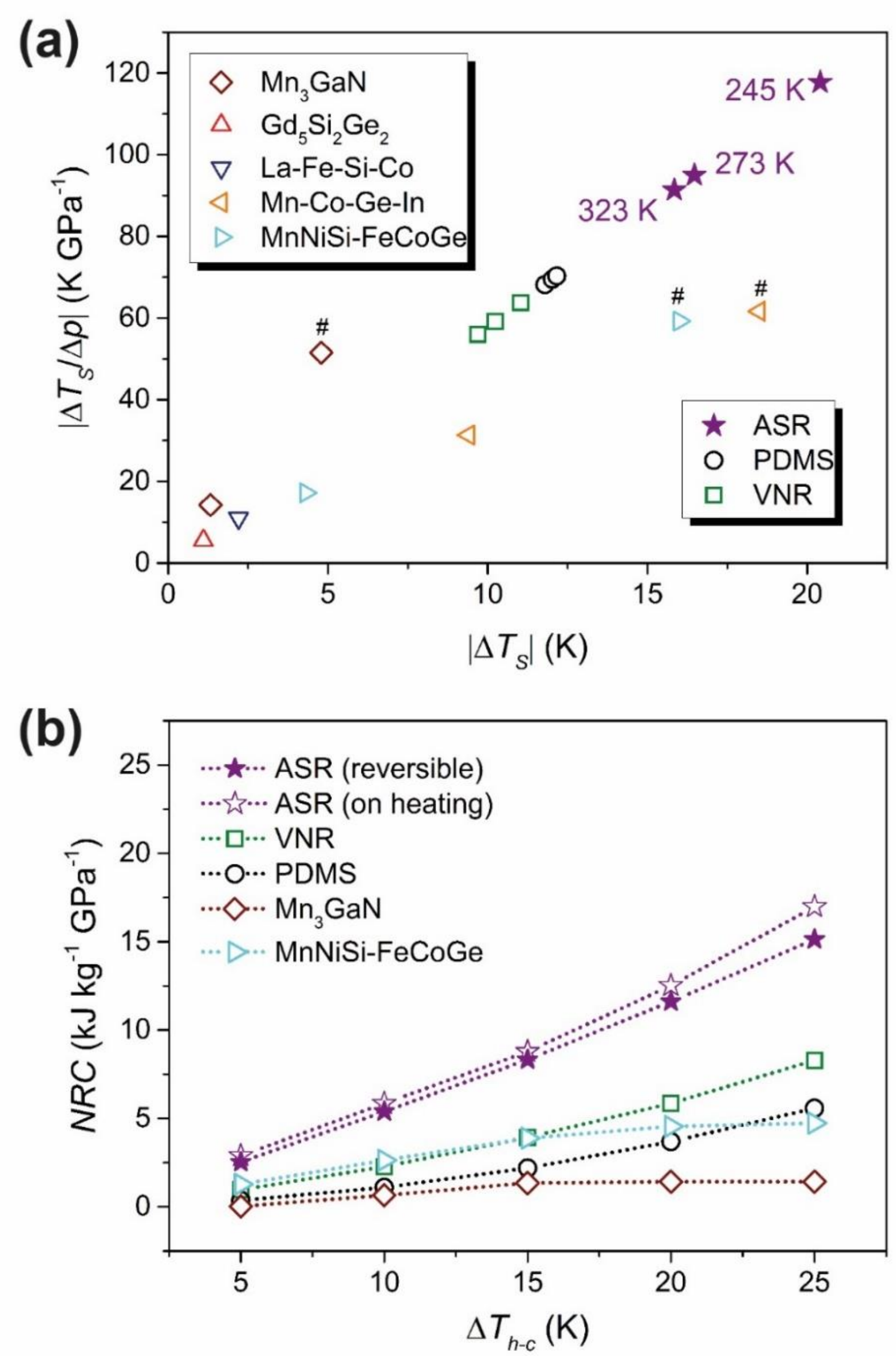

**Figure 4.** (Color online) Performance parameters. (a) Normalized adiabatic temperature change *vs.* temperature change for ASR ($|\Delta p|$ = 173 MPa), PDMS ($|\Delta p|$ = 173 MPa),[20] VNR ($|\Delta p|$ = 173 MPa),[19],[21] $Mn_3GaN$ (maximum $|\Delta \mathrm{T_S}|$ reported for $|\Delta p|$ = 93 MPa),[32] $Gd_5Si_2Ge_2$ (maximum $|\Delta T_S|$ reported for $|\Delta p|$ = 200 MPa),[33] La-Fe-Si-Co (maximum $|\Delta T_S|$ reported for $|\Delta p|$ = 200 MPa),[34] Mn-Co-Ge-In (maximum $|\Delta T_S|$ reported for $|\Delta p|$ = 300 MPa),[11] and MnNiSi-FeCoGe (maximum $|\Delta T_S|$ reported for $|\Delta p|$ = 250 MPa);[35] 245 K, 273 K and 323 K are selected temperatures for ASR; # represents data obtained from indirect methods. (b) Normalized refrigerant capacity *vs.* $\Delta T_{h\text{-}c}$ for ASR ($T_{hot}$ = 300 K and $|\Delta p|$ = 173 MPa; solid stars are related to reversible $\Delta S_T$ data and open stars are related to $\Delta S_T$ on heating), PDMS ($T_{hot}$ = 315 K and $|\Delta p|$ = 130 MPa), VNR ($T_{hot}$ = 315 K and $|\Delta p|$ = 173 MPa), $Mn_3GaN$ ($T_{hot}$ = 295 K and $|\Delta p|$ = 139 MPa), and MnNiSi-FeCoGe ($T_{hot}$ = 335 K and $|\Delta p|$ = 270 MPa); the dotted lines connecting the symbols are guides for the eyes.

## 4. CONCLUSIONS

Supergiant barocaloric changes are observed in ASR at moderate applied pressures and relatively low sample strains, reaching the maximum values of $|\Delta T_S|$ = 41.1 K (at ~298 K, for $|\Delta p|$ = 390 MPa) and reversible $|\Delta S_T|$ = 182(46) J kg$^{-1}$ K$^{-1}$ for $|\Delta p|$ = 173 MPa. The striking results for ASR also include a normalized adiabatic temperature change of ~120 K GPa$^{-1}$ for $|\Delta T_S|$ = 20.4 K and $|\Delta p|$ = 173 MPa, a normalized refrigerant capacity of ~15 kJ kg$^{-1}$ GPa$^{-1}$ for $\Delta T_{h\text{-}c}$ = 25 K and $|\Delta p|$ = 173 MPa, and a relative cooling power (*RCP*) of 13(3) kJ kg$^{-1}$ for $|\Delta p|$ = 173 MPa. The supergiant barocaloric values for ASR are due to combined effects of the first-order crystalline–amorphous transition and the polymer chain rearrangements unrelated to the crystallization process. These are the largest barocaloric effects known among elastomers so far. The major concern with elastomers is related to the low thermal conductivity, which can be improved by developing polymers with high conductivity fillers. But considering all the favorable characteristics exhibited by ASR concerning solid-state cooling, we can foresee a practical interest in the development of energy-efficient and environmental-friendly refrigeration devices based on barocaloric effect in polysiloxanes and other elastomers. As a further matter, pressure-induced crystallization on an amorphous polymer can dramatically modify its barocaloric properties; controlling or tuning the degree of crystallization during barocaloric processes may lead to even greater values of $\Delta T_S$, $\Delta S_T$, refrigerant capacity, and relative cooling power.

## ACKNOWLEDGMENT

The authors acknowledge financial support from FAPESP (project number 2016/22934-3), CNPq, CAPES and CNPEM.

NOMENCLATURE

| | |
|---|---|
| $\Delta S_T$ | isothermal entropy change |
| $\Delta T_S$ | adiabatic temperature change |
| $\sigma$-CE | mechanocaloric effect |
| $NR$ | natural rubber |
| $VNR$ | vulcanized natural rubber |
| $RTV\text{-}SR$ | room-temperature vulcanizing silicone rubbers |
| $ASR$ | acetoxy silicone rubber |
| $p$ | pressure |
| $\varepsilon$ | strain |
| $T$ | temperature |
| $l_{p,T}$ | length of the sample at pressure $p$ and temperature $T$ |
| $l_0$ | length of the sample at ambient pressure and room temperature |
| $A$ | cross-sectional area of the sample |
| $\Delta V$ | volume change |
| $XRD$ | X-ray diffraction |
| $FTIR$ | Fourier transform infrared spectroscopy |
| $T_{TR}$ | transition temperature |
| $\Delta p$ | pressure change |
| $T_{m,\Delta T}$ | temperature of the maximum $\Delta T$ |
| $T_{m,\Delta S}$ | temperature of the maximum $\Delta S$ |
| $\rho_0$ | density at ambient pressure and room temperature |
| $NRC$ | normalized refrigerant capacity |
| $\Delta T_{h-c}$ | temperature difference between hot and cold reservoirs |
| $T_{\text{hot}}$ | temperature of the hot reservoir |
| $T_{\text{cold}}$ | temperature of the cold reservoir |
| $RCP$ | relative cooling power |
| $\Delta S_{max}$ | maximum entropy change |
| $\delta T_{FWHM}$ | full width of peak at half maximum |

# Electronic Supplementary Information

# Supergiant Barocaloric Effects in Acetoxy Silicone Rubber over a Wide Temperature Range: Great Potential for Solid-state Cooling

William Imamura[a,], Érik Oda Usuda[b], Lucas Soares Paixão[c], Nicolau Molina Bom[d,e], Angelo Marcio Gomes[f], Alexandre Magnus Gomes Carvalho[g*]

[a] Faculdade de Engenharia Mecânica, UNICAMP, CEP 13083-860, Campinas, SP, Brazil.
[b] Universidade Federal de São Paulo, UNIFESP, CEP 09913-030, Diadema, SP, Brazil.
[c] Instituto de Física Gleb Wataghin, UNICAMP, CEP 13083-859, Campinas, SP, Brazil.
[d] Catalan Institute of Nanoscience and Nanotechnology, ICN2, 08860 Belaterra, Barcelona, Spain
[e] The Institute of Photonic Sciences, ICFO, 08193 Castelldefels, Barcelona, Spain
[f] Instituto de Física, UFRJ, CEP 21941-972, Rio de Janeiro, RJ, Brazil.
[g] Departamento de Engenharia Mecânica, UEM, 87020-900, Maringá, PR, Brazil.

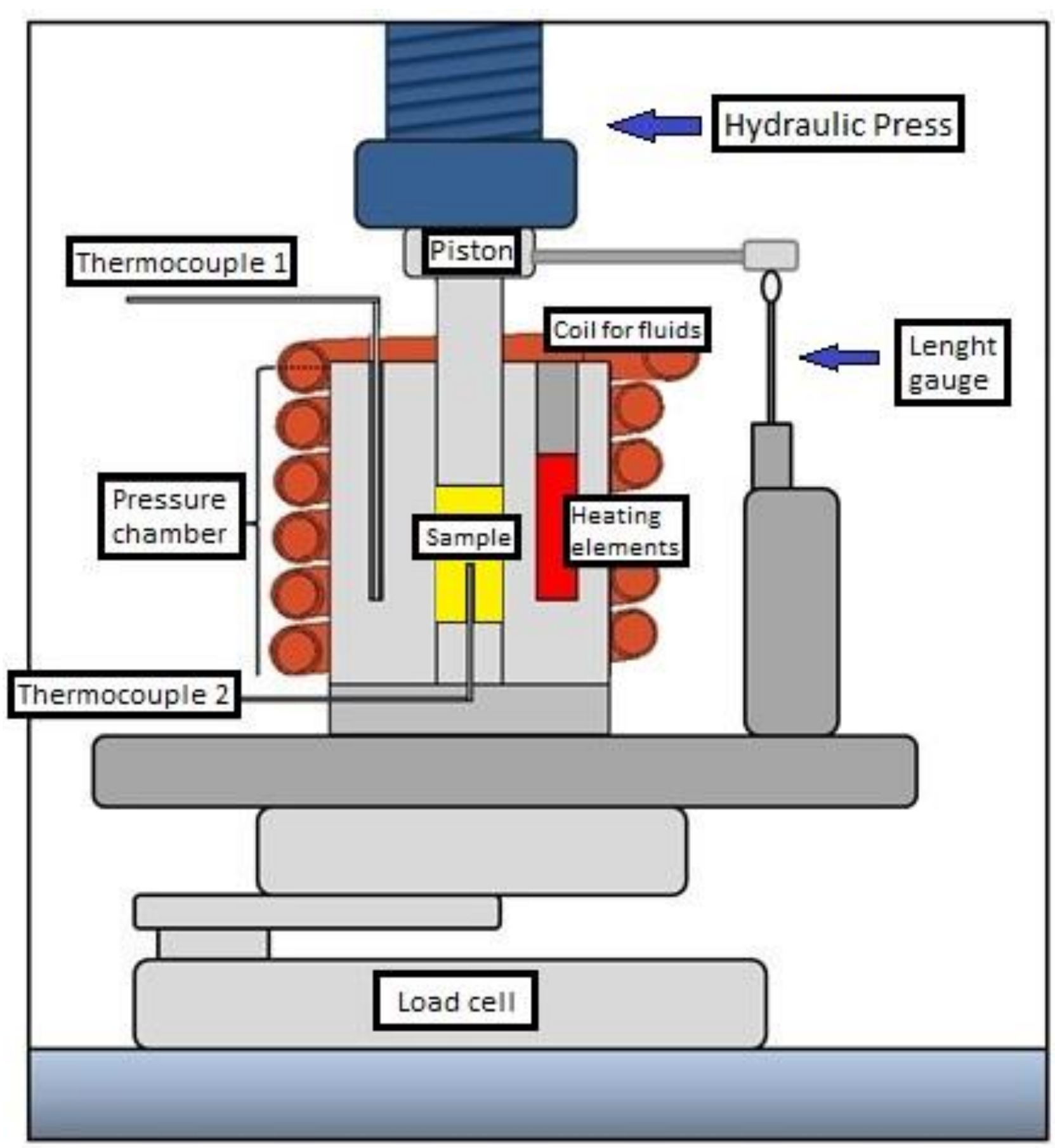

**Fig. S1** Diagram of the barocaloric apparatus.

**Table S1** List of equipment

| Equipment/Sensor | Brand/Model | Range | Accuracy |
|---|---|---|---|
| Manual Hydraulic Press | Bovenau P15000 | 0-15 tonf | - |
| Lengh Gauge | Heidenhein METRO ND 760 | 0-60 mm | 0.5 μm |
| Temperature controller + type K thermocouple | LakeShore Model 335 | 10-800 K | 0.1 K |
| Load cell | Alfa Instrumentos 3101C | 0-5 tonf | 0.5 kg |

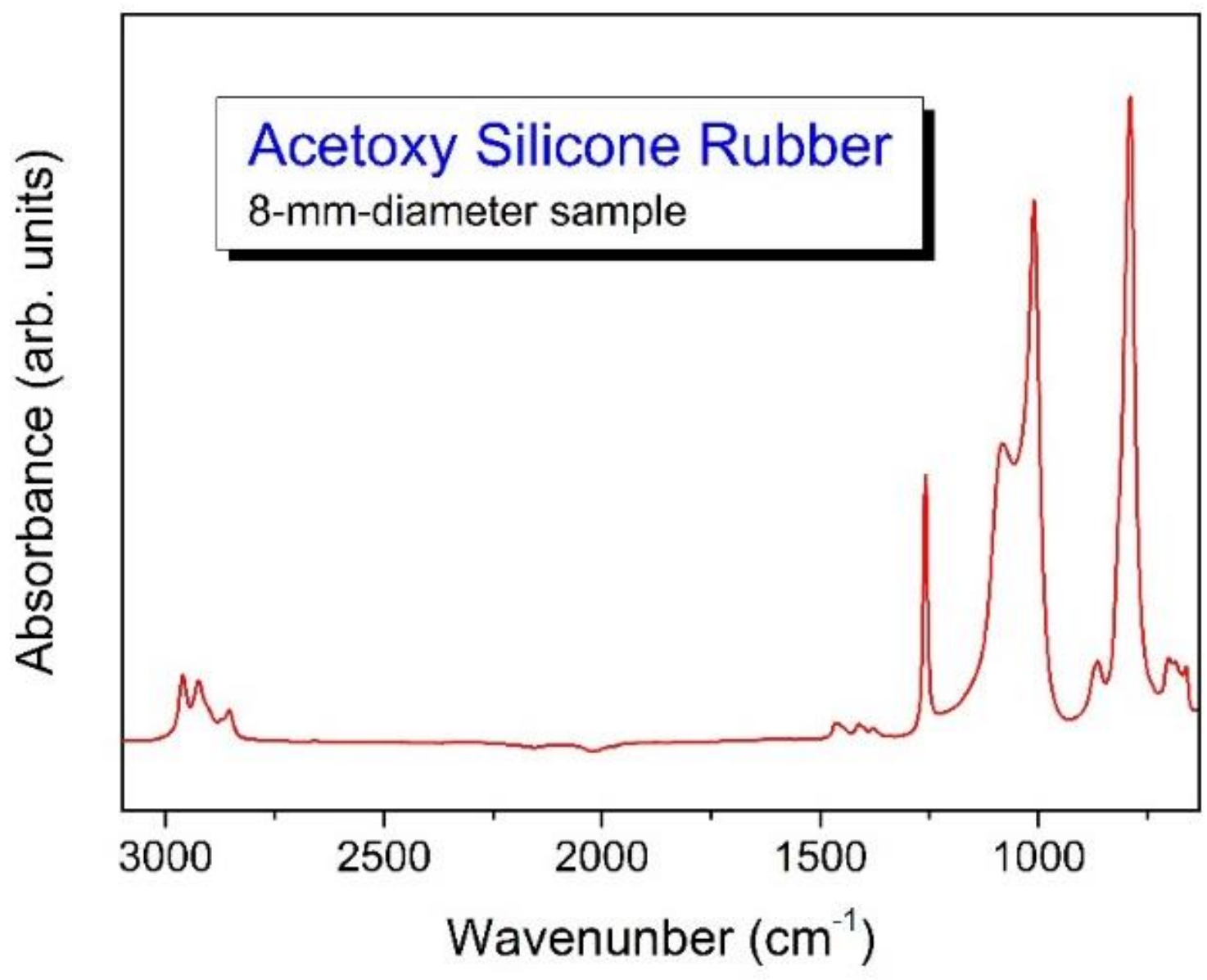

**Fig. S2** FTIR spectrum of ASR. The vibrational modes observed are typical of silicone rubber[1].

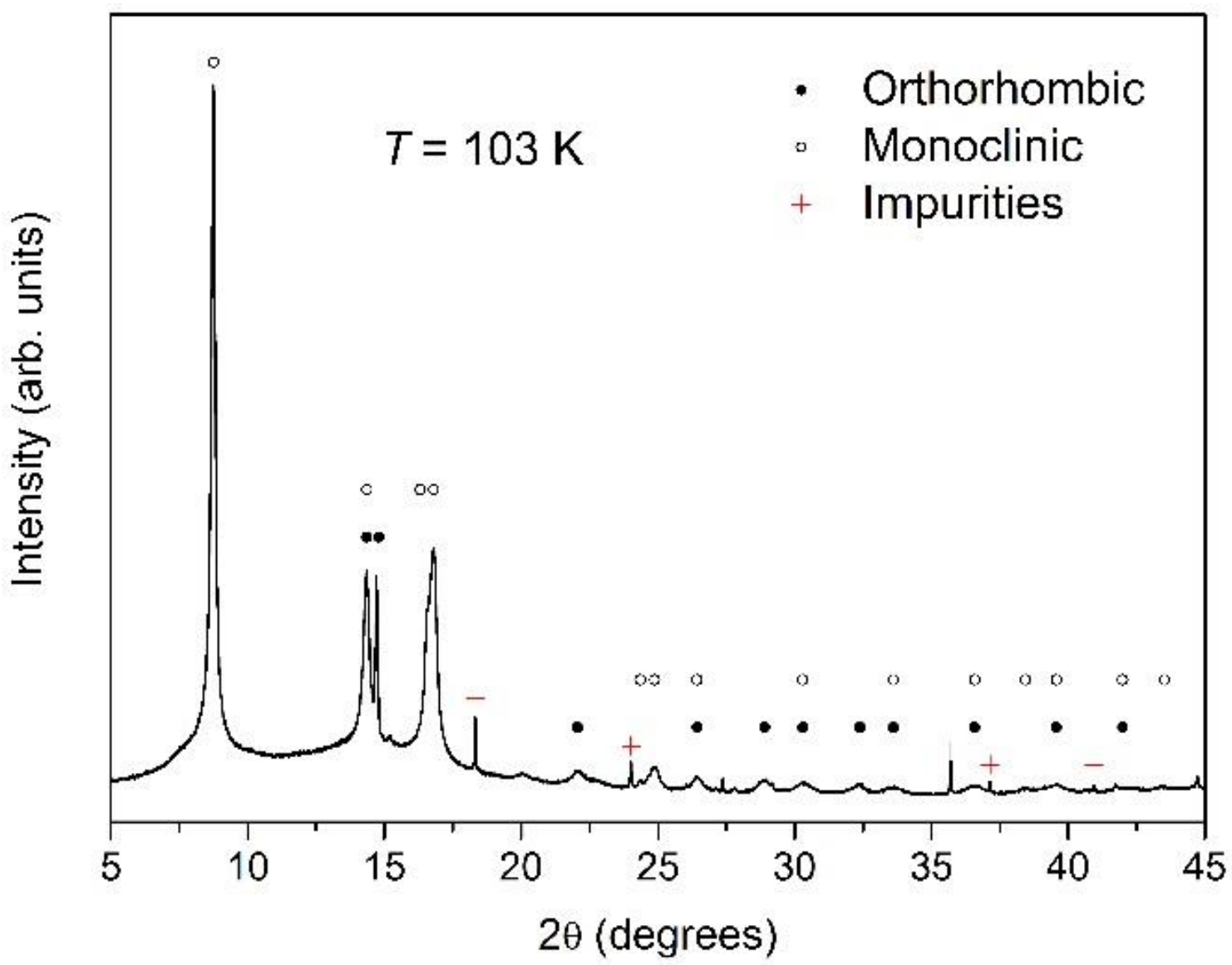

**Fig. S3** X-ray diffraction pattern of ASR obtained at 103 K and ambient pressure. See Table S-1 for crystallographic information.

**Table S2** Crystallographic parameters of ASR at 103 K and ambient pressure, based on the observed Bragg reflections.

| Phase | Space group | a [Å] | b [Å] | c [Å] | α [degree] | β [degree] | γ [degree] |
|---|---|---|---|---|---|---|---|
| Orthorhombic | Cmmm | 8.269 | 3.983 | 4.033 | 90 | 90 | 90 |
| Monoclinic | P2/m | 4.138 | 2.957 | 6.756 | 90 | 90.3 | 90 |